\documentclass[10pt,conference]{IEEEtran}
\usepackage{amsmath,amssymb,amsfonts}
\usepackage{color}
\usepackage{authblk}
\usepackage{multirow}
\usepackage{graphicx}
\usepackage{textcomp}
\usepackage[symbol]{footmisc}
\usepackage{url}
\usepackage{cite}
\usepackage{hyperref}

\newcommand{\ket}[1]{\ensuremath{\left|#1\right\rangle}}

\newcommand{\specialcell}[2][c]{\begin{tabular}[#1]{@{}l@{}}#2\end{tabular}}

\begin{document}
\title{A Scalable 5,6-Qubit Grover's Quantum Search Algorithm}
\author[1]{Dinesh Reddy Vemula}
\author[2]{Debanjan Konar}
\author[3]{Sudeep Satheesan}
\author[4]{Sri Mounica Kalidasu}
\author[2]{Attila Cangi}
\affil[1]{Department of Computer Science and Engineering, SRM University AP, Andhra Pradesh, India \authorcr Email: {\tt \{dineshvemula@gmail.com\}}}
\affil[2]{Center for Advanced Systems Understanding (CASUS) and Helmholtz-Zentrum Dresden-Rossendorf (HZDR), Germany \authorcr Email: {\tt \{d.konar@hzdr.de and a.cangi@hzdr.de\}}}
\affil[3] {Digital Lab \& COE, Firstsource Solutions Ltd, India  \authorcr Email: {\tt \{sudeeps78@gmail.com\}}} 
\affil[4]{Indian Institute of Technology Guwahati, India \authorcr Email: {\tt \{kalidasusm18@gmail.com\}}}

\maketitle
\begin{abstract}
 Recent studies have been spurred on by the promise of advanced quantum computing technology, which has led to the development of quantum computer simulations on classical hardware. Grover's quantum search algorithm is one of the well-known applications of quantum computing, enabling quantum computers to perform a database search (unsorted array) and quadratically outperform their classical counterparts in terms of time. Given the restricted access to database search for an oracle model (black-box), researchers have demonstrated various implementations of Grover's circuit for two to four qubits on various platforms. However, larger search spaces have not yet been explored. In this paper, a scalable Quantum Grover Search algorithm is introduced and implemented using 5-qubit and 6-qubit quantum circuits, along with a design pattern for ease of building an Oracle for a higher order of qubits. For our implementation, the probability of finding the correct entity is in the high nineties. The accuracy of the proposed 5-qubit and 6-qubit circuits is benchmarked against the state-of-the-art implementations for 3-qubit and 4-qubit. Furthermore, the reusability of the proposed quantum circuits using subroutines is also illustrated by the opportunity for large-scale implementation of quantum algorithms in the future.
\end{abstract}


\ifCLASSOPTIONpeerreview
\begin{center} \bfseries A Scalable 5,6-Qubit Grover's Quantum Search Algorithm \end{center}
\fi
%
\IEEEpeerreviewmaketitle

\section{Introduction}
\label{chapter:intro}

Quantum computing is a novel paradigm in computational science that has the potential to revolutionise the field~\cite{ladd2010}. Recently, quantum computing offers promising quantum advantages in various applications over classical counterparts due to the development of quantum hardware and inherent quantum entanglement characteristics~\cite{arute2019, cerezo2021}. On quantum computers, conventional NP-hard tasks~\cite{shor1997, grover1996, deutsch1992} can be solved in polynomial time, demonstrating a significant speedup over classical computers. Using Shor's method~\cite{shor1997} for factoring offers a super-polynomial speedup, whereas Grover's quantum search~\cite{grover1996} yields a polynomial speedup. The real-world applications of quantum computing and the speedups given by quantum algorithms will rely heavily on noisy intermediate-scale quantum (NISQ) devices~\cite{preskill2018}, which can handle hundreds of quantum bits. The behaviour and viability of quantum algorithms, as well as their future applications as quantum computing advances, have been studied in the last few decades~\cite{glos2022, ortiz2018}. However, the entire scope of classical issues that quantum computers can effectively handle has not yet been recognized~\cite{ortiz2018}. In addition to this, with the growing data size, the computational science community has a rising demand to successfully deploy quantum algorithms on large scale applications~\cite{wolf2017}. \\
Grover’s search algorithm is one of the most popular quantum computing algorithms that can search for an entity with a high probability in an unstructured list using only O($\sqrt{N}$)~\cite{bennett1997} evaluations. Grover's algorithm finds an entity in an unstructured list in fewer steps than a classical system~\cite{grover1997}. In quantum processing, the search entity is one of the states from among all possible superposition states. According to Grover's quantum search algorithm~\cite{grover1997}, there are ${2^n}$ = $N$ states in an $n$-qubit network, and the probability of finding every state is $\frac{1}{N}$. Hence, the amplitude of each state is $\frac{1}{\sqrt{N}}$. The same problem takes maximum O($N$)~\cite{grover1997} trials in the classical system. According to Bennett \emph{et al.}~\cite{bennett1997}, no quantum method can solve the database search problem in fewer than O($\sqrt{N}$) steps asymptotically. Later, Boyer \emph{et al.}~\cite{boyer1999} have shown that there are no quantum algorithm can outperform Grover's algorithm by more than a few percentage points with a $50\%$ probability of success. The applications of Grover's search include solving collision problems, finding the mean and median~\cite{grover1997} of a given dataset, and reverse-engineering cryptographic hash functions by allowing an attacker to find a victim's password. \\
In the last few decades, a plethora of Grover's quantum search algorithms have been explored, which includes theoretical proof of the efficiency of the algorithm~\cite{zalka1999} and its applications in various problem domains~\cite{durr1998, brassard1998, hoyer1998, cerf2000, mahmud2022}. Of late, the experimental implementation of fast Grover's quantum search algorithm has been extended to Nuclear Magnetic Resonance (NMR)~\cite{chuang1998} for a system of four states. Moreover, a generalized Grover's search algorithm~\cite{boyer1999} is introduced to deal with more than one marked state and allow an arbitrary number of unitary transformations in the original quantum circuit~\cite{grover1998}. It may be noted that the generalization of Grover's quantum search algorithm leads to adiabatic quantum computing settings~\cite{roland2002}. However, the NMR systems lack scalability and optimization during the presentation of their experimental results. Grover’s quantum search algorithm has been implemented for two to four qubits on IBM's quantum computer, such as a 2-qubit implementation on the \emph{ibmqx2} architecture~\cite{gurnani2021}, a 3-qubit on a trapped-ion architecture~\cite{figgatt2017}, and a 4-qubit circuit on \emph{ibmqx5} architecture~\cite{stromberg2018}. There is an implementation of Grover's search on a 2-qubit, 4-state quantum computer by Brickman \emph{et al.}~\cite{brickman2005}, and their findings demonstrate an improvement over the corresponding classical search approach. Mandviwalla \emph{et al.}~\cite{mandiwalla2018} demonstrate the experiments of a single-pattern Grover's search for up to four qubits on IBM's quantum computer and present the accuracy and execution time results. Recently, a five-qubit Grover's search is realised on the \emph{ibmqx4} quantum computer by Abhijith \emph{et al.}~\cite{abhijith2022}, and their models suffer from a lower success rate. Moreover, FPGA-based emulation of Grover's circuit was suggested by Lee \emph{et al.}~\cite{lee2016} and Mahmud \emph{et al.}~\cite{mahmud2022}; however, their hardware designs are also not scalable. \\
Recent years have witnessed the growing interest in hybrid quantum-classical variational algorithms to solve the scalability issues of Grover's quantum search~\cite{morales12018}. Using Grover's quantum search algorithm as a test case for variational hybrid quantum-classical algorithms has been proven to be asymptotically optimal. However, the variational quantum algorithms are restricted to four qubits. It uses only one shared angle in the quantum settings, which is not applicable for five or more qubits. The Grover search circuit proposed in our study has been modified from the usual circuit and widened to allow dynamic single-pattern and multi-pattern searches. The number of entangled qubits simulated in our experiments is also larger than in earlier studies. The significant contribution of this article is as follows:
\begin{enumerate}
    \item This paper introduces a $C^4Z$ gate and a novel Oracle function for 5-qubit and 6-qubit circuits using the $C^4Z$ gate. Hence, a scalable Quantum Grover Search algorithm is proposed.  
    \item Furthermore, we developed a V-shaped Oracle (V- Oracle) pattern, which uses fewer gates and would potentially lead to fewer gate errors. The V-Oracle architecture can be easily extended to a higher number of qubits. The diffusion function takes the form \emph{HX + Oracle + XH} to increase the amplitude of the search state.
    \item Finally, the experiments presented here are performed on the IBM Q experience, with each circuit being iterated $1024$ times. The accuracy of the proposed 5-qubit and 6-qubit circuits is benchmarked against the state-of-the art implementations for 3-qubit and 4-qubit.
\end{enumerate}
The rest of this article's sections are arranged in the following fashion. Section~\ref{grover_search:algorithm} explains the proposed novel scalable Grover's quantum search algorithm based on $5$ and $6$ qubits circuits, including an introduction to the gate construction. Grover's quantum search algorithms in $5$ and $6$ qubit space and experimental results are provided in Section~\ref{results:implementation}. Section~\ref{addendum} sheds light on the scalability and reusability of the proposed Grover's quantum search model. Finally, the concluding remarks and future research directions with the opportunity for large-scale implementation of Grover's quantum search are confabulated in Section~\ref{conclud}. The complete circuit to implement 5 and 6 -qubit Grover quantum search is provided in \emph{Appendix}.
\section{Gate Construction and Proposed Scalable Grover's Quantum Search Algorithm} 
\label{grover_search:algorithm}

Searching an unstructured data array of $N$ items is the primary objective of Grover's quantum search algorithm~\cite{grover1996}. For Grover's quantum search strategy to be successful, it must locate an element in the collection $A=A_1,A_2,A_3,\ldots S_N$ where $N$ is the number of elements in the set $A$. The Boolean function $\delta$ must be able to find an element in the collection so that $\delta(x) \rightarrow[0, 1]$~\cite{mahmud2022}. When iterating over each element one at a time, a classical computer would need an average of $\frac{N}{2}$ queries for an array of cardinality $N$~\cite{williams2011}. A quadratic speedup over conventional computers is achieved by utilising Grover's technique on a quantum machine, which only requires $\sqrt{N}$ queries. Multiple entries in the same collection may be found using Grover's quantum search technique~\cite{boyer1999}. Each of the target patterns is found with equal probability by the algorithm. An Hadamard gate is first applied to transform the input qubits into entanglement and superposition with equal coefficients. Phase inversion and inversion around the mean are performed on the initial quantum state~\cite{mahmud2022}. The following gates are required for implementing Grover’s quantum search algorithm.\\

\textit{Controlled $T$-gate ($cT$)}: 
A T gate is a phase shift gate that shifts the phase of a qubit by $\frac{\pi}{4}$ around the Z-axis. The $cT$ or controlled T gate, is a two-qubit gate created by adding a control qubit to a $T$ gate. For example, if the control qubit equals to $|1\rangle$, a $T$ gate will be applied to the target qubit; otherwise, it will stay unaltered. The generic $c^kT$ gate, where $k$ is the number of control qubits, may be created by adding more control qubits. A controlled $T$-gate describes a rotation of $\frac{\pi}{4}$ around the Z-axis on the target qubit when the control qubit is true. ${T^{\dagger}}$, its conjugate transpose, describes a rotation of $-\frac{\pi}{4}$ around the same axis. The matrix representations of controlled $T$ gates can be found in~\cite{figgatt2017, mahmud2022}, as shown below.
\begin{equation} 
T =\begin{bmatrix} 
1 & 0\\ 0 & e^{i\frac{\pi}{4}}
\end{bmatrix},\;\; 
cT =\begin{bmatrix} 
1 & 0 & 0 & 0\\ 0 & 1 & 0 & 0\\ 0 & 0 & 1 & 0\\ 0 & 0 & 0 & e^{i\frac{\pi}{4}} 
\end{bmatrix}.
\end{equation} 

\textit{Controlled $Z$-gate ($cZ$)}:
If it needs to invert the phase of the input qubit (i.e., the $|1\rangle$ basis state) while keeping the $|0\rangle$ basis state unaltered, we use a $Z$ gate (or Phase Inversion gate)~\cite{williams2011}. A controlled $Z$-gate describes a rotation of ${\pi}$ around the Z-axis on the target qubit when the control qubit is true. It is possible to generate a $c^kZ$ gate by adding several control qubits that are all equal to $|1\rangle$. When this occurs, a $Z$ gate is applied to the target qubit, which otherwise remains unaffected. We have used $cZ$ gate in our proposed model available in IBM Q. A $Z$ and $cZ$ gate matrix representation is shown below~\cite{figgatt2017, mahmud2022}.
\begin{equation*} 
Z=\begin{bmatrix}
$1$ & $0$\\ $0$ & $-1$ 
\end{bmatrix},\;\;\; 
cZ=\begin{bmatrix}
$1$ & $0$ & $0$ & $0$\\ $0$ & $1$ & $0$ & $0$\\ $0$ & $0$ & $1$ & $0$\\ $0$ & $0$ & $0$ & $-1$ 
\end{bmatrix}.
\end{equation*} 
A quantum computer can solve a type of decision problem called the bounded-error quantum polynomial time (BQP) problem in polynomial time with an error probability of at most $\frac{1}{3}$ in all cases~\cite{cerf2000}. The unstructured search problem in this instance becomes a BQP problem as Grover’s iteration, when applied O($\sqrt{N}$) times, finds the element with a probability greater than $\frac{2}{3}$~\cite{nielsen2001, bernstein1997}.\\
The Grover iteration, or Grover operator, is a quantum circuit that is employed in the quantum search process. There are two phases in the Grover iteration~\cite{grover1996}. At the outset, the oracle function flips the phase of a single amplitude in a marked state. Once the indicated state amplitude has been flipped over, the diffusion layer has completed its task. While the other states retain their original amplitude, the target state is in an inverted condition, causing the amplitude of the target state to rise significantly and the amplitude of the other states to fall somewhat. 
\subsection{Oracle}
The oracle~\cite{williams2011} is frequently portrayed as a black box that takes the input state, $|x\rangle$, and inverts the select coefficient of the target base state.\\
\textit{Initialization}: 
The Hadamard gate ($H$ gate) is used in the first stage of the algorithm to place all of the qubits in superposition~\cite{williams2011} due to its ability to produce an equal superposition of basic states. If an $H$ gate is applied to the $|0\rangle$ state, the resultant state is an equal probability superposition of the $|0\rangle$ and $|1\rangle$ states, \emph{i.e.},$\frac{1}{\sqrt{2}}(|0\rangle+|1\rangle)$. The following matrix represents an $H$ gate:
\begin{equation}
    H = \frac{1}{\sqrt{2}} \begin{bmatrix}1 & \phantom{-}1\\ 1 & -1 \end{bmatrix}.
\end{equation}
The oracle function performs a phase flip on the marked state. Initially, the amplitude of all states, including the marked one, are $\frac{1}{\sqrt{N}}$. The mean amplitude of all states is also $\frac{1}{\sqrt{N}}$. The phase flip transforms the amplitude of the marked state ($\alpha^{(0)}$) to  $-\frac{1}{\sqrt{N}}$. It changes the mean amplitude as follows.
\begin{eqnarray} 
\mu^{(0)} =  \frac{((N-1)*a^{(0)} + \alpha^{(0)})}{N}\quad \\=  \frac{(( (N-1)*\frac{1}{\sqrt{N}}-\frac{1}{\sqrt{N}}))}{N} =\frac{(N-2)}{(N\sqrt{N})} 
\end{eqnarray} 
Where $a$ is the amplitude of the unmarked state and $\alpha$ is the amplitude of the marked state. The superscript $0$ refers to time zero, or, when the first iteration of Oracle is applied. We present two models for Oracle function here: one is the ${C^{4}}Z$ gate model and the other is the V-Oracle model.

\subsubsection{Quaternary controlled Pauli Z-gate ($C^4Z$) model} 
\begin{figure*}[htbp] 
\centering
  \includegraphics[scale=0.45]{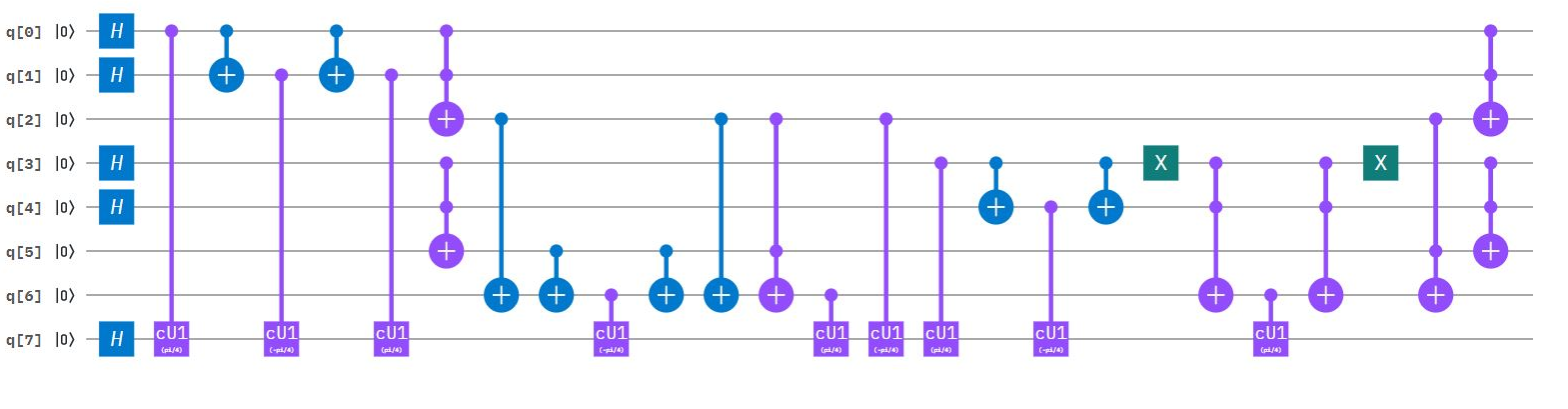} 
  \caption{$C^4$Z Model: Quantum circuit for a 5-qubit search space using $T$ and ${T^{\dagger}}$ gates. The sequence of gates correspond to a phase shift of $\pi$ when the input is $|11111\rangle$}
  \label{image-cZModelfor5} 
 \end{figure*}

$C^2Z$ and $C^3Z$ gates are used for the implementation of 3-qubit and 4-qubit quantum circuits~\cite{stromberg2018}, respectively. These gates are constructed using \emph{CNOTs} and $T$-gates~\cite{takahashi2014}. We have built a $C^4$Z composite gate to implement Oracle for a search space of five qubits, as shown in Figure~\ref{image-cZModelfor5}. This Oracle is used for search state $\ket{11111}$. There are $2^5 = 32$ states for five qubits. For all states except $\ket{11111}$, the output will be the same as the input without any phase change. When the state is $\ket{11111}$, all five $T$-gates are activated and the third ${T^\dagger}$ gate also gets activated, resulting in a phase shift of {$\pi$}. The state of qubit-4 changes from $\ket{1}$ to $e^{i\pi}$ $\ket{1}$, which is ($\cos \pi + i \sin \pi$) $\ket{1}$ = -$\ket{1}$. Therefore, $\ket{11111}$  becomes -$\ket{11111}$. The $cU1$s are used to replicate $T$ and ${T^{\dagger}}$ gates. If we apply the gate logic for all other states, it can be validated that $T$ and ${T^{\dagger}}$ gate(s) cancel out each other, and the output state will be same as the input state without any phase transformation.

\subsubsection{V-Oracle model}
\begin{figure}[htbp]
 \centering
 \includegraphics[scale=0.5]{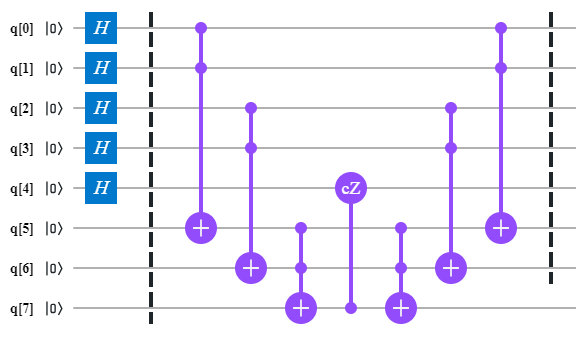} 
 \caption{V-Oracle - a symmetric structure with Toffoli, $cZ$-gate and ancilla qubits for 5-Qubits}
 \label{image-Voraclefor5Qubits}
 \end{figure}

\begin{figure}[htp]
 \centering
 \includegraphics[scale=0.5]{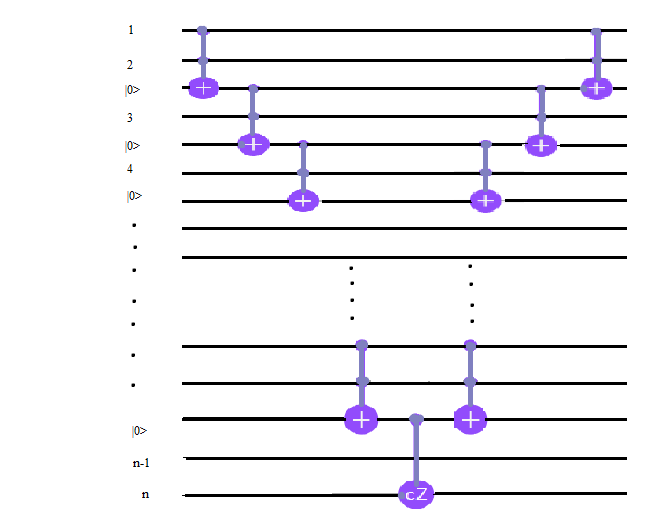} 
 \caption{Scaling V-Oracle for an even number of qubits}
 \label{image-VoraclefornevenQubits} 
 \end{figure}

\begin{figure}[htp]
 \centering
 \includegraphics[scale=0.5]{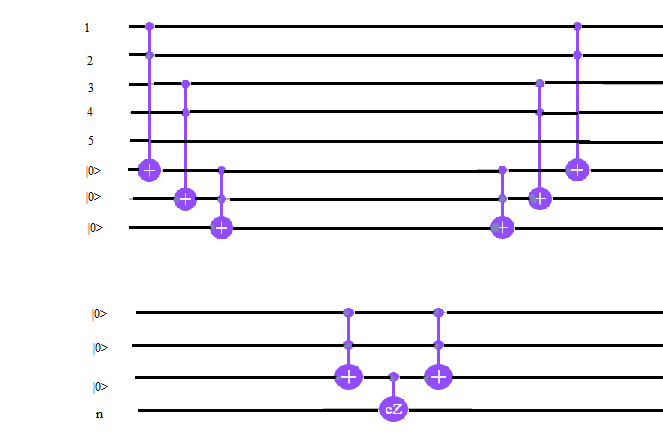} 
 \caption{Scaling V-Oracle for an odd number of qubits} 
 \label{image-VoraclefornoddQubits} 
 \end{figure}

With an increase in the number of qubits, the implementation of the ${C^{n-1}}Z$ gate becomes more complex. It requires more number of gates to implement this ${C^{n-1}}Z$ gate, thus potentially increasing gate error. In addition, there is a lack of a standard pattern to build this ${C^{n-1}}Z$ gate. V-Oracle has a symmetric structure designed using a combination of \emph{Toffoli}, $cZ$ gate and \emph{ancilla} qubits, as shown in Figure~\ref{image-Voraclefor5Qubits}. This pattern is extensible to a higher number of qubits, and hence this V-Oracle model offers the scalability of Grover's quantum search space. It provides a standard way to design an oracle for a larger number of qubits. Furthermore, this implementation requires fewer gates. Figure~\ref{image-VoraclefornevenQubits} and Figure~\ref{image-VoraclefornoddQubits} show the generic V-Oracle model for an even and an odd number of qubits. The $\ket{0}$ in the figure represents \emph{ancilla} qubits. V-Oracle for n-qubit can be designed using ${2*(n-2)}$ \emph{Toffoli} gates, one $cZ$ gate and $(n-2)$ \emph{ancilla} qubits.

\subsubsection{Gate error}
Each quantum state is affected by a small error introduced by quantum gates. Qubits are uniquely entangled with one another because of the hardware implementation and layout of the coupling map. Depending on the qubit, the error size may vary. The V-Oracle employs a fewer number of gates compared to the ${C^{n-1}}Z$-model, thus resulting in smaller gate error. We adopted this V-Oracle pattern in the rest of our experimentation as it generates a shallower overall circuit as compared to the ${C^{n-1}}Z$ model.

\subsection{Diffusion}
\begin{figure}[htbp] 
\centering
\includegraphics[height=2.0in, width =3.5in]{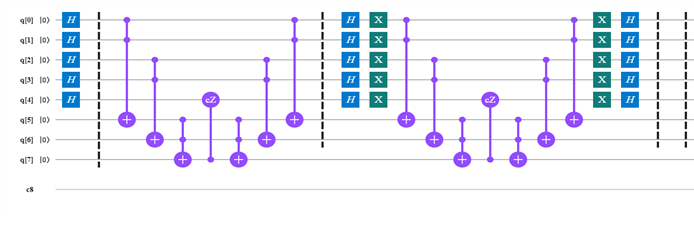} 
\caption{ Proposed implementation of quantum circuit for 5-qubit search space using the combination of Superposition followed two iterations of Oracle followed by Diffusion.} 
\label{image-SOD5qubit}
\end{figure}

Diffusion can be implemented as \emph{HX + Oracle + XH} as seen in Figure~\ref{image-SOD5qubit}, where $H$ is the Hadamard transform and $X$ is the $Pauli-X$ gate. This combination inverts the amplitudes around their mean. Grover's approach predicts an increase in the marked state's amplitude of $\frac{1}{\sqrt{N}}$. The formula for the new amplitude of the marked state $\alpha^{(1)}$ is 2*$\mu^{(0)}$- $\alpha^{(0)}$. With $\alpha^{(0)}$= -$\frac{1}{\sqrt{N}}$ and $\mu^{(0)}$=  $\frac{(N-2)}{(N\sqrt{N})}$ (from equation 3), we can derive $\alpha^{(1)}$ as follows:
\begin{eqnarray} 
\alpha^{(1)}=2*\frac{(N-2)}{(N\sqrt{N})} -\frac{1}{\sqrt{N}} \\= \frac{1}{\sqrt{N}}*(2*\frac{(N-2)}{(N)}+1)
= \frac{3}{\sqrt{N}}- \frac{4}{N\sqrt{N}}  
\end{eqnarray}
For a 5-qubit search space, $N = {2^n} = 32$. Hence, one iteration of Oracle and Diffusion will yield amplitude = $0.53033 - 0.02209$ or equals to $0.50824$ and a probability of the marked state is ${(0.50824)^{2}}$ or equals to $25.83\%$.
Similarly, for a 6-qubit space, one iteration would yield a probability of $13.48\%$. Table~\ref{table-table1} compares the derived probabilities with simulating probabilities using the V-Oracle design. The deviations are observed to be very less, except for the 6-qubit circuit.
\begin{table}[t]
\centering
\caption{Derived vs simulating probabilities after one iteration of Oracle and Diffusion} 
\begin{tabular}{|p{30pt}|p{70pt}|p{70pt}|}
\hline
Search Space & \specialcell{Derived probability \\using the Equation 6}  & \specialcell{simulating probability\\ (V-Oracle)-1024 shots} \\
\hline
3-qubit & $78.12\%$ & $78.125\%$ \\
\hline
4-qubit & $47.26\%$ & $47.07\%$  \\
\hline
5-qubit & $25.83\%$ & $25.195\%$  \\
\hline
6-qubit & $13.48\%$ & $14.94\%$  \\
\hline
\end{tabular}
\label{table-table1}
\end{table}
Proceeding with the 5-qubit quantum circuit, after the second rotation, the amplitude of a marked state $\alpha^{(2)}$ is 2$*\mu^{(1)}$- $\alpha^{(1)}$, where $\alpha^{(1)}$ is $0.50824$ (for 5-qubit). To find $\mu^{(1)}$, we need the amplitudes of all states, excluding marked states. The sum of the probabilities for unmarked ($31$) states = $1$ - probability of marked state (square of the amplitude given in Equation 6).
\begin{eqnarray}
\text{Sum of probabilities for unmarked states}\\ =  1 - {(\frac{3}{\sqrt{N}}-  \frac{4}{N\sqrt{N}})}^2 \\
\text{Probability of each unmarked state}\\ = \frac{1- {(\frac{3}{\sqrt{N}}-  \frac{4}{N\sqrt{N}})}^2}{N-1} \\
\text{Amplitude of each unmarked state}\\ =  \sqrt{ \frac{1- {(\frac{3}{\sqrt{N}}-  \frac{4}{N\sqrt{N}})}^2}{N-1}} \\
 =  \frac{1}{N\sqrt{N}}*\sqrt{(\frac{({N}^3-9{N}^2+24N+16)}{(N-1)})} \\ 
= \frac{1}{N\sqrt{N}}*{\sqrt{\frac{(N-1)*({N}^2-8N+16)}{(N-1)}}} \\
= \frac{1}{N\sqrt{N}}* {\sqrt{{(N-4)}^2}}  =  \frac{(N-4)}{(N\sqrt{N})} 
\end{eqnarray}
Hence, the amplitude of each unmarked state after the first rotation is $\frac{(32-4)}{(32\sqrt{32})}$ or equals to $0.15467$.\\
Mean, $\mu^{(1)}$ = $\frac{(31*0.15467-0.50824)}{32}=0.13395$ \\
So, $\alpha^{(1)} = 2*0.13395-(-0.50824) = 0.77614$ \\
The probability of a marked state after the second iteration will be ${(0.77614)^2}$ or equals to $60.24\%$. \\
We derived the amplitudes of marked states for up to four rotations. Table \ref{table-table2} provides corresponding probabilities along with simulating outputs. Table \ref{table-table3} presents similar results for the 6-qubit circuit for up to six rotations.
\begin{table}[t]
\centering
\caption{Derived vs simulating probabilities for 5-qubit search space} 
\begin{tabular}{|p{60pt}|p{70pt}|p{70pt}|}
\hline
Search Space after & \specialcell{Derived probability \\using the Equation 6}  & \specialcell{simulating probability\\ (V-Oracle)-1024 shots} \\
\hline
First rotation & $25.83\%$ & $25.195\%$ \\
\hline
Second rotation & $60.24\%$ & $61.13\%$  \\
\hline
Third rotation & $89.69\%$ & $89.94\%$  \\
\hline
Fourth rotation & $99.92\%$ & $99.9\%$  \\
\hline
\end{tabular}
\label{table-table2}
\end{table}
\begin{table}[t]
\centering
\caption{Derived vs simulating probabilities for 6-qubit search space} 
\begin{tabular}{|p{60pt}|p{70pt}|p{70pt}|}
\hline
Search Space after & \specialcell{Derived probability \\using the Equation 6}  & \specialcell{simulating probability\\ (V-Oracle)-1024 shots} \\
\hline
First rotation & $13.48\%$ & $14.258\%$ \\
\hline
Second rotation & $34.39\%$ & $34.863\%$  \\
\hline
Third rotation & $59.14\%$ & $60.254\%$  \\
\hline
Fourth rotation & $81.64\%$ & $82.227\%$  \\
\hline
Fifth rotation & $96.35\%$ & $97.168\%$  \\
\hline
Sixth rotation & $99.66\%$ & $99.902\%$  \\
\hline
\end{tabular}
\label{table-table3}
\end{table}
It has been observed from the experimental results that for 5 and 6 qubit implementations, the derived and simulating outputs are almost similar.
\subsubsection{Unstructured search as a BQP~\cite{nielsen2001} problem}
\begin{figure}[htbp] 
\centering
\includegraphics[height=2.0in, width =3.5in]{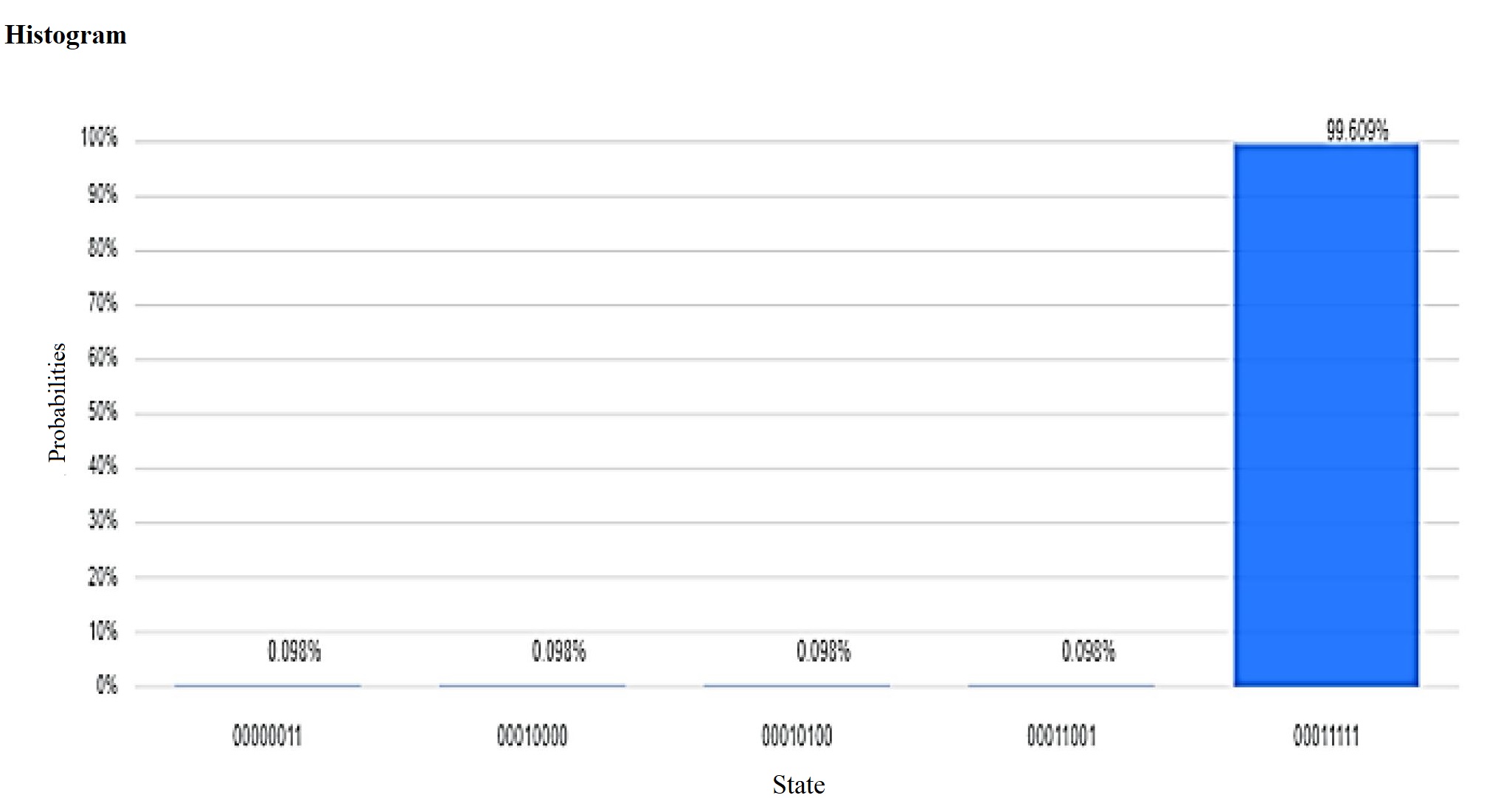} 
\caption{V-Oracle implementation for 5-qubit showing high probability of $11111$ state.} 
\label{image-5v11111}
\end{figure}

With $\sqrt{N}$ number of rotations, the probability of the marked state increases above $\frac{2}{3}$. To attain a probability of at least $\frac{2}{3}\sim 0.6667$, the amplitude required for the marked state would be greater than or equal to $\sqrt{0.6667} \geqslant 0.8165$. Hence, the number of repetitions of the order is greater than or equal to $0.8165\sqrt{N}$. This is owing to the fact that with each rotation (Oracle and Diffusion), the amplitude increases by at least $\frac{1}{\sqrt{N}}$, \emph{i.e.}, $\geqslant \frac{1}{\sqrt{N}}$. Therefore, with $0.8165\sqrt{N}$ rotations, amplitude will be greater than or equal to $(0.8165\sqrt{N})*\frac{1}{\sqrt{N}}\geqslant 0.8165 $. It translates to a probability greater than or equal to $0.8165^2 \geqslant 0.6667$, as desired. For a 5-qubit space, this would evaluate to $0.8165\sqrt{32}$ or equals to $4.62 \sim 5$. While the probability with five qubits is $\geqslant \frac{2}{3}$, we observed that with four rotations the probability was maximum, in the high $90$s, as given in Figure~\ref{image-5v11111}.

\section{Results} 
\label{results:implementation}

This section presents the experimental results of Grover's quantum search implementation in 5-qubit and 6- qubit space on the IBM Q simulator. By default, IBM Q had five qubits and we extended it by editing the code in the IBM Q composer to increase the number of qubits. \emph{E.g.}, increase the size of the variable \emph{qreg} from ``$q[4]$'' to ``$q[8]$'' to add four more qubits in the circuit (the full circuit is presented in \emph{Appendix}).

\subsection{Grover’s implementation in 5-qubit space}
The core part of the Oracle remains the same for all the states. The idea is to invert the qubit(s) using the $Pauli-X$ gate to flip the amplitude of the desired state. \emph{E.g.}, to design an Oracle for state $01110$, we apply the Oracle from the earlier section and flip the first and last qubits. It will convert from $-11111$ to $-01110$.
\begin{figure}[htbp] 
\centering
\includegraphics[height=2.0in, width =3.5in]{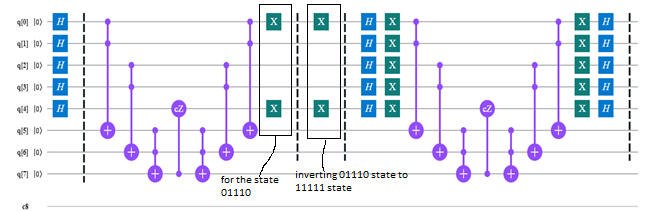}
\label{image-Figure6a} 
\includegraphics[height=2.0in, width =3.5in]{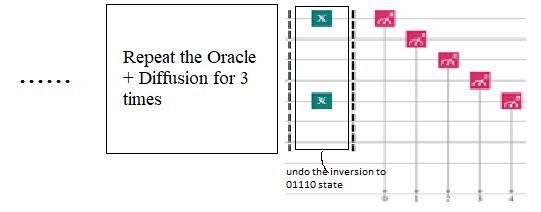} 
\caption{Grover’s algorithm implementation of 5-qubit search space for the input $01110$ using the repeated combination of Oracle followed by Diffusion four times.} 
\label{image-5v01110} 
\end{figure}
\begin{figure}[htbp] 
\centering
\includegraphics[height=2.0in, width =3.5in]{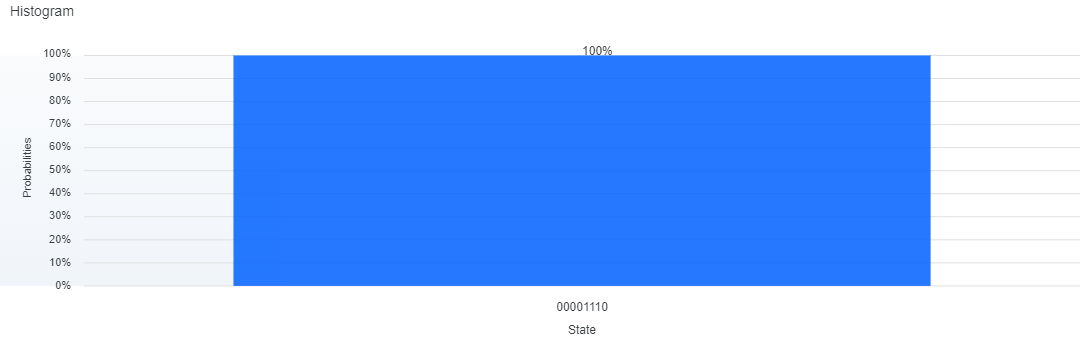} 
\caption{Result for 5-qubit showing 100\% probability for $01110$ state} 
\label{image-5v01110graph} 
\end{figure}
The circuit for Grover’s quantum search implementation of the $01110$ states is shown in Figure~\ref{image-5v01110}, and the circuit has additional $X$ gates to invert the phase of the search state $01110$. There are two layers of $X$-gates, one for inversion immediately after the oracle, other at the end just before measurement. These layers are required to amplify the probability of the search state. These $X$-gates are applied only to qubits in $\ket{0}$ state, \emph{e.g.}, for the search state$01110$, inversion and undo-inversion are applied only to the first and last qubits at the two layers $X$-gates, respectively. It may be noted that the combination of Oracle and Diffusion is repeated four times. The graph as shown in Figure~\ref{image-5v01110graph} yields $100\%$ probability for the state of $01110$ on the IBM Q simulator.

\subsection{Grover’s Implementation for 6-qubit space}
The Oracle and diffusion for 6-qubit are shown in Figure~\ref{image-SOD6qubit} and Figure~\ref{image-SOD6qubitresult}. The complete circuit is given in \emph{Appendix}.
\begin{figure}[htbp] 
\centering
\includegraphics[height=2.0in, width =3.5in]{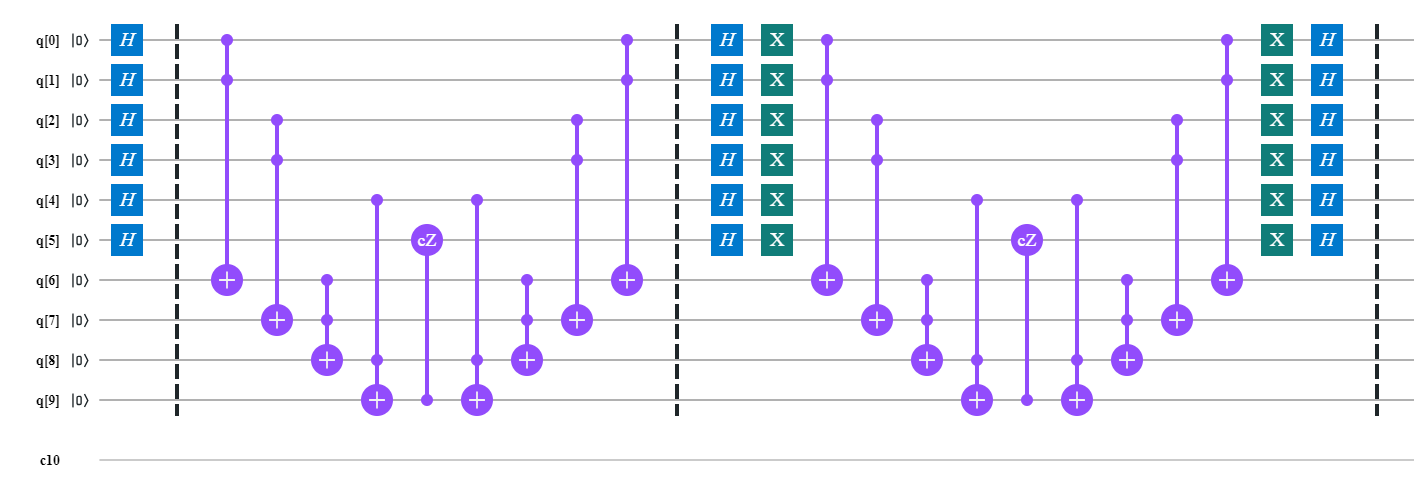} 
\caption{This diagram represents the proposed architecture for a 6-qubit search space with the help of four ancilla qubits. It depicts superposition followed by two iterations of Oracle followed by Diffusion.} 
\label{image-SOD6qubit} 
\end{figure}
\begin{figure}[htbp] 
\centering
\includegraphics[height=2.0in, width =3.5in]{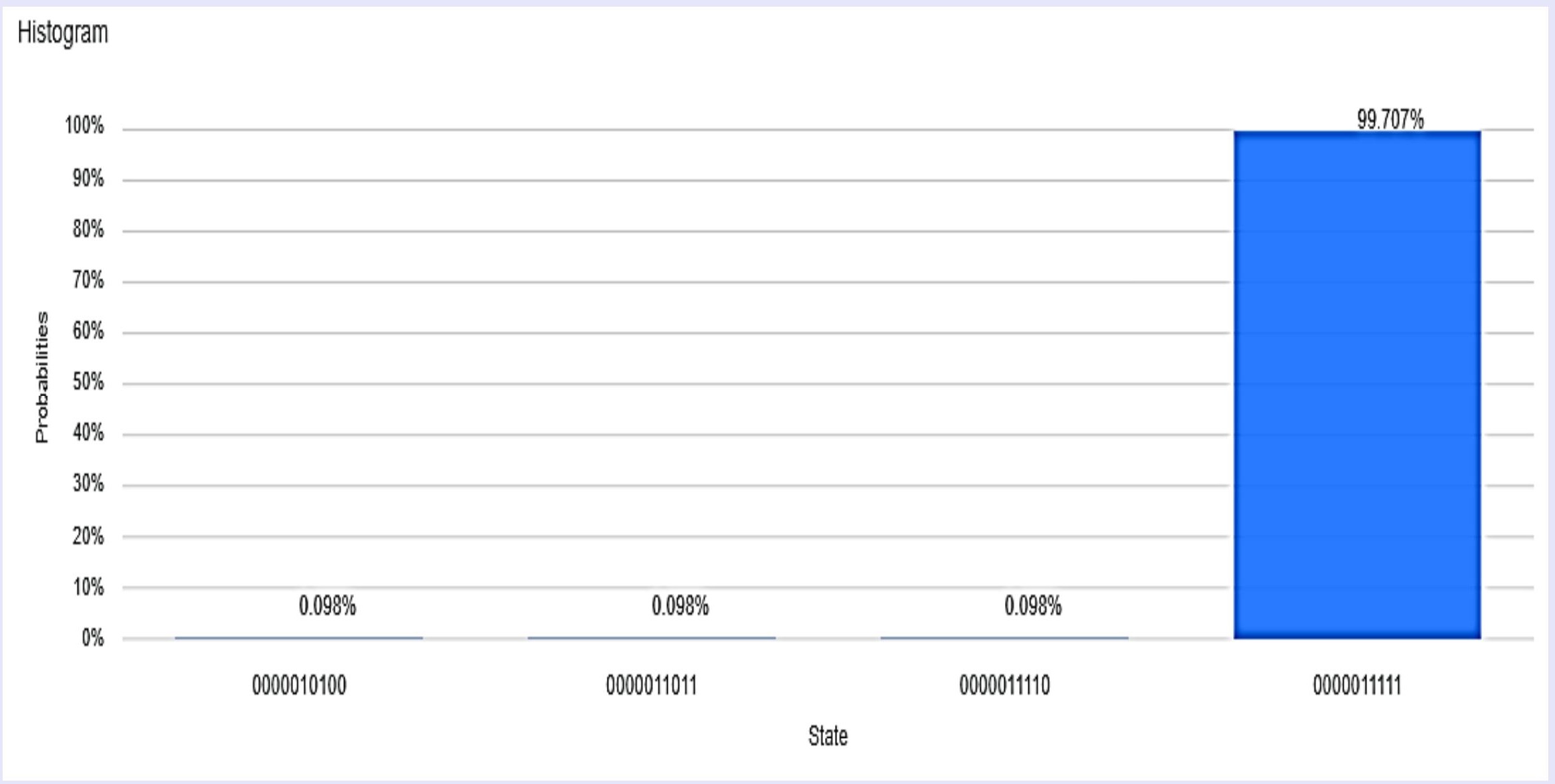}
\caption{Experimental result for 6-qubit showing a high probability of $111111$ state.} 
\label{image-SOD6qubitresult} 
\end{figure}
Along similar lines, oracle and diffusion for higher qubits can be easily implemented using the V-Oracle model. Table~\ref{table-table4} reports the probability of finding a given state using the proposed $C^4Z$ and V-Oracle model for the $2$ to $3$-qubit quantum circuit and the $4$ to $6$ qubit quantum circuit, respectively, with the state-of-the art techniques implemented on the IBM Q simulator with $2$ to $4$-qubit quantum circuits~\cite{gurnani2021, mandiwalla2018, stromberg2018}. For the 4-qubit quantum circuit, the proposed approach offers a higher probability owing to the fact that the Oracle followed by Diffusion is repeated thrice. It may be noted that similar experimental results are reported for fewer qubits on the IBM Q platform. 
\begin{table}[t]
\centering
\caption{Comparative experimental results in terms of simulating probability} 
\begin{tabular}{|p{20pt}|p{40pt}|p{60pt}|p{60pt}|}
\hline
Number of Qubits & Number of rotations (Oracle and Diffusion)  & Simulating probability using the proposed quantum circuit & Simulating probability using the state-of-the art techniques\\
\hline
$2$ & $1$ & $100\%$ & $100\%$~\cite{gurnani2021} \\
\hline
$3$ & $2$ &  $95\%$  &  $95\%$~\cite{mandiwalla2018}\\
\hline
$4$ & $3$ & $96.387\%$ & $47\%$~\cite{stromberg2018} \\
\hline
$5$ & $4$ & $\sim100\%$  &  ~ \\
\hline
$6$ & $5$  &  $99.805\%$  &  ~ \\
\hline
\end{tabular}
\label{table-table4}
\end{table}

\section{Discussion}
\label{addendum}
\subsection{Reusability}

\begin{figure}[htbp] 
\centering
\includegraphics[height=2.0in, width =3.5in]{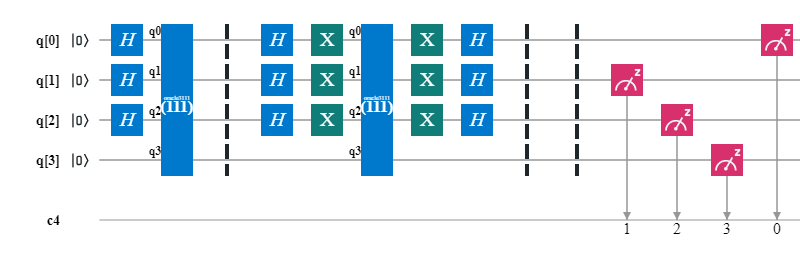} 
\caption{3-qubit search circuit using subroutines} \label{image-3qubitsubroutines} 
\end{figure}

The Grover’s rotation (Oracle and Diffusion) when repeated $0.8165\sqrt{N}$ times, increases the probability of finding the marked state to greater than $\frac{2}{3}$. For a 5-qubit space, this would evaluate to $0.8165\sqrt{32} =  4.62 \sim 5$. Since diffusion contains Oracle (\emph{HX + Oracle + XH}), the Oracle circuit is repeated ten times and it makes the quantum circuit complex and inefficient.\\
However, if the Oracle is built as a subroutine, we gain all the advantages of reusability. This reduces manual errors and it is less tedious as it avoids repeating all the gates and IBM Q provides subroutines. Despite the fact that the feature is not yet functional, it allows us to design the circuit. Figure~\ref{image-3qubitsubroutines}  illustrates the 3-qubit search circuit using subroutines with one rotation (Oracle and Diffusion). 

\subsection{Observations}

This subsection covers a few observations that can be handy while designing circuits on the IBM Q.
\subsubsection{Phase detection} 
\begin{figure}[htbp] 
    \centering
	\includegraphics[height=2.0in, width =3.5in]{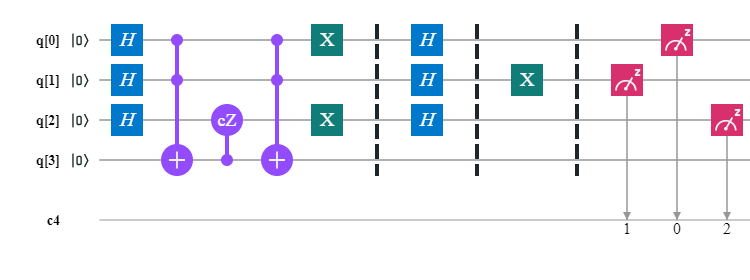}
	\caption{Circuit for 010 phase detection} 
	\label{image-010phasecircuit} 
	\includegraphics[height=2.0in, width =3.5in]{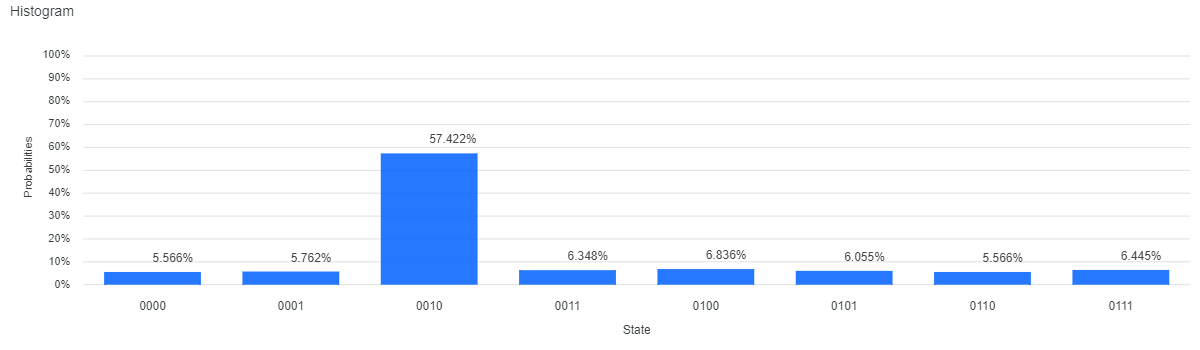} 
	\caption{Result for 010 phase detection} 
	\label{image-010phaseresult} 
\end{figure}

The IBM Q simulator yields  probabilities as an outcome after simulating a quantum circuit. Hence, it is a daunting task to figure out the correctness of an Oracle circuit as the output probability will always be positive. In order to validate whether a search state has its phase reversed or not, a circuit, as shown in Figure~\ref{image-010phasecircuit}, can be built by applying $H$ gates to all qubits between the first and second barriers. Between second and third barriers, $X$ gates are placed only for qubits that have $\ket{1}$ in the search state. For example, if a search state is $010$, place $X$ gate between the second and third barriers only for the second qubit. If the search state is $000$, then do not place any $X$ gates; for $111$, keep $X$ for all three qubits. The output will show all the states; however, if the phase of the search state is negative, as is the case for a correct Oracle, there will be a distinct tower for that state, as shown in Figure~\ref{image-010phaseresult}. If the Oracle is incorrect, the search state will be lost among other states, \emph{i.e.}, it will be as tall as other states.\\
For controlled phase shift on the IBM Q simulator, $cRz$ does not yield accurate results, unlike $cU1$. A 5-qubit Oracle and Diffusion with $cRz$ ($-\frac{\pi}{4}$) resulted in a probability of approximately $14\%$ after one cycle, whereas the same circuit with $cU1$ gives approximately $25\%$ theoretically. The same result is observed with the existing 4-qubit quantum circuit as well. It is worth noting that the $cA$ combination of \emph{HX + Oracle + XH} usually works as a diffusion

\section{Conclusion}
\label{conclud}

This paper presents a 5-qubit and 6-qubit implementations of Grover’s quantum search algorithm. It introduces a V-Oracle pattern that offers a standard and relatively feasible technique to implement an unstructured search for a higher number of qubits. When applied for fewer qubits (less than $5$), the V-Oracle model gives equivalent or outperforms the state-of-the-art approaches. All the experiments have been performed on the IBM Q simulator. The experiments suggest that future quantum systems can use the proposed scalable Grover's search technique and technology for various large-scale applications. However, the \url{ibmq\_16\_melbourne} backend results are not promising, which indicates that the actual qubits still suffer from a lack of coherence. The authors are currently working in this direction.

\section{Acknowledgements}

This work was partially supported by the Center for Advanced Systems Understanding (CASUS), financed by Germany’s Federal Ministry of Education and Research (BMBF) and by the Saxon state government out of the State budget approved by the Saxon State Parliament. We also appreciate the usage of IBM Q in this project. The views expressed here belong entirely to the authors and not the IBM Q experience team or any other entity. 
\section*{Appendix}
\label{Appendix}
%
\begin{figure}[htbp] 
\centering
\includegraphics[height=4.0in, width =6.5in]{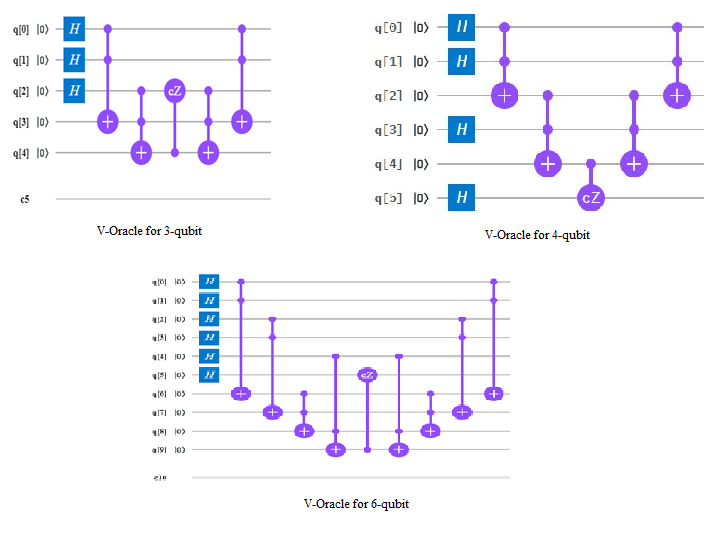}
\caption{V-Oracle for 3, 4 and 6 qubits search space.}
\label{image-Oracles3to6} 
\end{figure}
\begin{figure*}[htbp] 
\centering
\includegraphics[height=9.0in, width =6.5in]{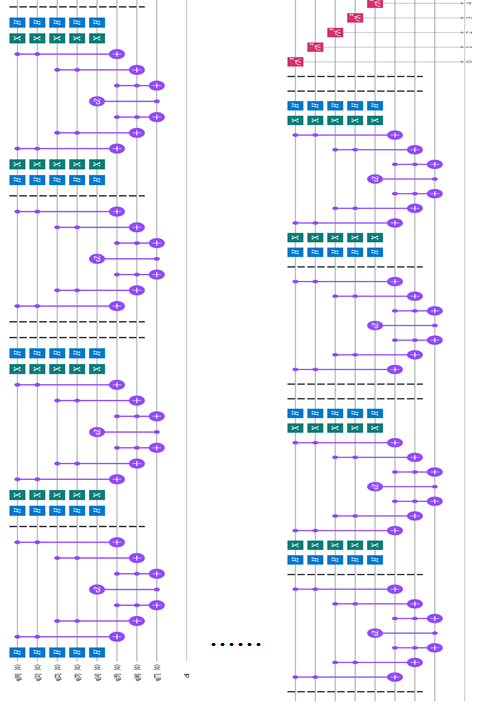}
\caption{Complete circuit for 5-qubit Grover’s implementation with $4$ rotations (Oracle followed by Diffusion) that yielded a probability of more than $99\%$. The diffusion is of the form \emph{HX + Oracle + XH}, hence we see the Oracle getting repeated $8$ times.}
\label{image-5qubitcircuit} 
\end{figure*}
\begin{figure*}[htbp] 
\centering
\includegraphics[height=9.0in, width =6.5in]{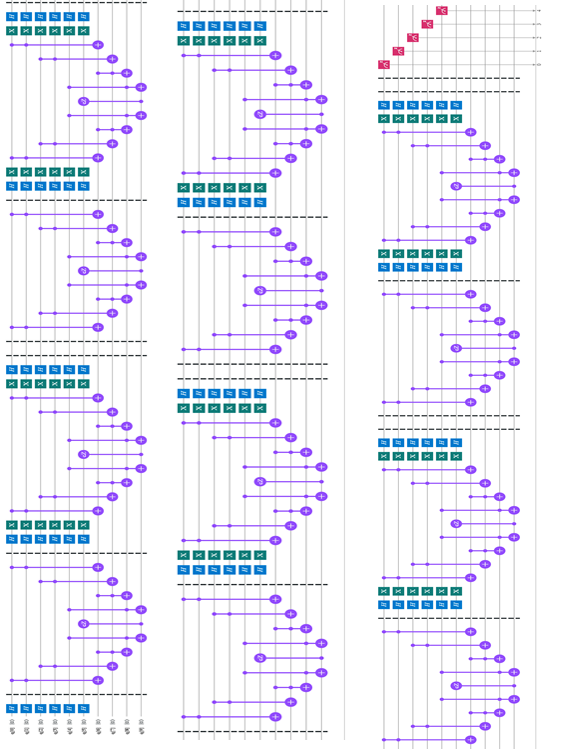}
\caption{Complete circuit for 6-qubit Grover’s implementation with $6$ rotations (Oracle followed by Diffusion) that yielded a probability of more than $99\%$. The diffusion is of the form HX + Oracle + XH, hence we see the Oracle getting repeated $12$ times.}
\label{image-6qubitcircuit} 
\end{figure*}
\clearpage

\begin{thebibliography}{1}
\bibitem{ladd2010}
T.~D.~Ladd, F.~Jelezko, R.~Laflamme, \emph{et al.}, ``Quantum computers,” \emph{Nature}, vol. 464, no. 7285, p. 45--53, 2010, doi: \url{https://doi.org/10.1038/nature08812}.
\bibitem{arute2019}
F.~Arute, K.~Arya, and R.~Babbush, \emph{et al.}, ``Quantum supremacy using a programmable superconducting processor,” \emph{Nature}, vol. 574, pp. 505-–510, 2019, doi: \url{https://doi.org/10.1038/s41586-019-1666-5}.
\bibitem{cerezo2021}
M.~Cerezo, A.~Arrasmith, and R.~Babbush, \emph{et al.}, ``Variational quantum algorithms,” \emph{Nat. Rev. Phys.}, vol. 3, pp. 625-–644, 2021, doi: \url{https://doi.org/10.1038/s42254-021-00348-9}.
\bibitem{shor1997}
P.~W.~Shor, ``Polynomial-time algorithms for prime factorization and discrete logarithms on a quantum computer," \emph{SIAM J. Comput.}, vol. 26, no. 5, pp. 1484--1509, 1997, doi: \url{https://doi.org/10.1137/S0097539795293172}.
\bibitem{grover1996}
L.~K.~Grover, ``A fast quantum mechanical algorithm for database search," \emph{In Proc. 28th Annu. ACM Symp. Theory Comput.}, pp. 212--219, 1996, doi: \url{https://doi.org/10.1145/237814.237866}.
\bibitem{deutsch1992}
D.~Deutsch and R.~Jozsa, ``Rapid solution of problems by quantum computation," \emph{In Proc. Roy. Soc. London A}, vol. 439, no. 1907, pp. 553--558, 1992, doi: \url{https://doi.org/10.1098/rspa.1992.0167}.
\bibitem{preskill2018}
J.~Preskill, ``Quantum computing in the NISQ era and beyond," \emph{Quantum}, vol. 2, pp. 79, 2018, doi: \url{https://doi.org/10.22331/q-2018-08-06-79}.
\bibitem{glos2022}
A.~Glos, A.~Krawiec, and Z.~Zimborás, ``VariSpace-efficient binary optimization for variational quantum computing," \emph{npj Quantum Inf.}, vol. 8, no. 39, 2022, doi: \url{https://doi.org/10.1038/s41534-022-00546-y}.
\bibitem{ortiz2018}
A.~Perdomo-Ortiz,  M.~Benedetti, J.~Realpe-Gómez and R.~Biswas, ``Opportunities and challenges for quantum-assisted machine learning in near-term quantum computers", \emph{Quantum Sci. Technol.}, vol. 3, no. 030502, 2018, \url{https://doi.org/10.1088/2058-9565/aab859}.
\bibitem{wolf2017}
R.~de Wolf, ``The potential impact of quantum computers on society," \emph{Ethics Inf. Technol.}, vol. 19, pp. 271-–276, 2017, doi: \url{https://doi.org/10.1007/s10676-017-9439-z}.
\bibitem{bennett1997}
C.~H.~Bennett, E.~Bernstein, G.~Brassard, and U.~Vazirani, ``Strengths and Weaknesses of Quantum Computing," \emph{SIAM Journal on Computing}, vol. 26, no. 5, pp. 1510-1523, 1997, doi:\url{10.1137/S0097539796300933}.
\bibitem{grover1997}
L~K.~Grover, ``Quantum mechanics helps in searching for a needle in a haystack," \emph{Physical review letters}, vol. 79, no. 2, pp. 325, 1997, doi:\url{https://doi.org/10.1103/PhysRevLett.79.325}.
\bibitem{boyer1999}
M.~Boyer, G.~Brassard, P.~Hoyer, and A.~Tapp, ``Tight Bounds on Quantum Searching," \emph{Fortschr. Phys.}, vol. 46, no. 4/5, pp. 493--505, 1998, doi: \url{https://onlinelibrary.wiley.com/doi/10.1002/(SICI)1521-3978(199806)46:4/5\%3C493::AID-PROP493\%3E3.0.CO;2-P}.
\bibitem{zalka1999}
C.~Zalka, ``Grover's quantum searching algorithm is optimal," \emph{Phys. Rev. A}, vol. 60, no. 4, pp. 2746--2751, 1999, doi: \url{10.1103/PhysRevA.60.2746}.
\bibitem{durr1998}
C.~Durr, and P.~Hoyer, ``A Quantum Algorithm for Finding the Minimum," doi: \url{https://doi.org/10.48550/arXiv.quant-ph/9607014}, 1998.
\bibitem{brassard1998}
G.~Brassard, P.~Hoyer, and A.~Tapp, ``Quantum cryptanalysis of hash and claw-free functions," \emph{In proc. LATIN 1998: Theoretical Informatics, Springer, Berlin, Heidelberg}, pp. 163–169, vol 1380, 1998, doi: \url{https://doi.org/10.1007/BFb0054319}.
\bibitem{hoyer1998}
G.~Brassard, P.~Hoyer, and A.~Tapp, ``Quantum counting," \emph{In proc. Automata, Languages and Programming: ICALP 1998, Springer, Berlin, Heidelberg}, vol 1443, 1998, doi: \url{https://doi.org/10.1007/BFb0055105}.
\bibitem{cerf2000}
N.~J.~Cerf, L.~K.~Grover, and C.~P.~Williams, ``Nested quantum search and structured problems," \emph{Phys. Rev. A}, vol. 61, pp. 032303, 2000, doi: \url{https://doi.org/10.1103/PhysRevA.61.032303}.
\bibitem{mahmud2022}
N.~Mahmud, B.~Haase-Divine, A.~MacGillivray and E.~El-Araby, ``Quantum Dimension Reduction for Pattern Recognition in High-Resolution Spatio-Spectral Data," \emph{IEEE Transactions on Computers}, vol. 71, no. 1, pp. 1--12, 2022, doi: \url{10.1109/TC.2020.3034883}.
\bibitem{chuang1998}
I.~L.~Chuang, N.~Gershenfeld, and M.~Kubinec, ``Experimental Implementation of Fast Quantum Searching," \emph{Phys. Rev. Lett.}, vol. 80, pp. 3408, 1998, doi: \url{https://doi.org/10.1103/PhysRevA.61.032303}.
\bibitem{grover1998}
L.~K.~Grover, ``Quantum Computers Can Search Rapidly by Using Almost Any Transformation," \emph{Phys. Rev. Lett.}, vol. 80, pp. 4329, 1998, doi: \url{https://doi.org/10.1103/PhysRevLett.80.4329}.
\bibitem{roland2002}
J.~Roland, and N.~J. ~Cerf, ``Quantum search by local adiabatic evolution," \emph{Phys. Rev. A}, vol. 65, pp. 042308, 2002, doi: \url{https://doi.org/10.1103/PhysRevA.65.042308}.
\bibitem{gurnani2021}
K.~Gurnani, K.~B.~Behera, and P.~K.~Panigrahi, ``Demonstration of optimal fixed-point quantum search algorithm in IBM quantum computer," 2021, \url{https://doi.org/10.48550/arXiv.1712.10231}.
\bibitem{figgatt2017}
C.~Figgatt, D.~Maslov, A.~K.~Landsman, N.~M.~Linke, S.~Debnath, and C.~Monroe, ``Complete 3-Qubit Grover search on a programmable quantum computer," \emph{Nat. Commun.}, vol. 8, no. 1918, 2017. doi: \url{https://doi.org/10.1038/s41467-017-01904-7}.
\bibitem{stromberg2018}
P.~Str{\"o}mberg, and V.~B.~Karlsson, ``4-qubit Grover's algorithm implemented for the ibmqx5 architecture," \emph{Dissertation-KTH Royal Institute}, 2018, \url{https://www.diva-portal.org/smash/get/diva2:1214481/FULLTEXT01.pdf}.
\bibitem{brickman2005}
K.~A.~Brickman, P.~C.~Haljan, P.~J. Lee, M.~Acton, L.~Deslauriers, and C.~Monroe, ``Implementation of Grover's quantum search algorithm in a scalable system," \emph{Phys. Rev. A}, vol. 72, no. 5, Nov. 2005, doi: \url{https://doi.org/10.1103/PhysRevA.72.050306}.
\bibitem{mandiwalla2018}
A.~Mandviwalla, K.~Ohshiro and B.~Ji, ``Implementing Grover's algorithm on the IBM quantum computers," \emph{In. Proc. IEEE Int. Conf. Big Data}, pp. 2531--2537, 2018, doi: \url{10.1109/BigData.2018.8622457}.
\bibitem{abhijith2022}
J.~Abhijith \emph{et al.}, ``Quantum algorithm implementations for beginners," \emph{ACM Transactions on Quantum Computing}, 2022, doi: \url{https://doi.org/10.1145/3517340}.
\bibitem{lee2016}
Y.~H.~Lee, M.~Khalil-Hani, and M.~N.~Marson, ``An FPGA-based quantum computing emulation framework based on serial-parallel architecture," \emph{Int. J. Reconfigurable Comput.}, vol. 2016, 2016, doi: \url{https://doi.org/10.1155/2016/5718124}.
\bibitem{morales12018}
M.~E.~S.~Morales, T.~Tlyachev, and J.~Biamonte, ``Variational learning of Grover’s quantum search algorithm," \emph{Phys. Rev. A}, vol.98, pp. 062333, 2018, doi: \url{https://doi.org/10.1103/PhysRevA.98.062333}.
\bibitem{williams2011}
 C.~P.~Williams, ``Explorations in Quantum Computing," \emph{Texts in Computer Science, Springer}, 2011, doi: \url{https://doi.org/10.1007/978-1-84628-887-6}.
 \bibitem{nielsen2001}
 A.~M.~Nielsen and L.~I.~Chuang, ``Quantum computation and quantum information," \emph{Phys. Today}, vol. 54, pp. 60--72, 2001, 
\bibitem{bernstein1997}
E.~Bernstein, and U.~Vazirani, ``Quantum complexity theory," \emph{SIAM Journal on computing}, vol. 26, no. 5, pp. 1411--1473, 1997, doi: \url{https://doi.org/10.1137/S0097539796300921}.
\bibitem{takahashi2014}
K.~Takahashi, M.~Kurokawa, and M.~Hashimoto, ``Multi-layer quantum neural network controller trained by real-coded genetic algorithm," \emph{Neurocomputing}, vol. 134, pp. 159–-164, 2014, doi: \url{https://doi.org/10.1016/j.neucom.2012.12.073}










%

%





\end{thebibliography}

\end{document}